# Composition variations in Cu(In,Ga)(S,Se)₂ solar cells: not a gradient, but an interlaced network of two phases

Aubin JC. M. Prot[1,a)], Michele Melchiorre[1], Felix Dingwell[1], Anastasia Zelenina[2], Hossam Elanzeery[2], Alberto Lomuscio[2], Thomas Dalibor[2], Maxim Guc[3], Robert Fonoll-Rubio[3], Victor Izquierdo-Roca[3], Gunnar Kusch[4], Rachel A. Oliver[4], Susanne Siebentritt[1]

## AFFILIATIONS

[1]Laboratory for Photovoltaics, Physics and Materials Science Research Unit, University of Luxembourg, 41 rue du Brill, Belvaux L-4422, Luxembourg

[2]AVANCIS GmbH, Otto-Hahn-Ring 6, 81739 München, Germany

[3]Catalonia Institute for Energy Research (IREC), Jardins de les Dones de Negre 1, 2ª pl., 08930 Sant Adrià de Besòs, Barcelona, Spain

[4]Department of Materials Science and Metallurgy, University of Cambridge, 27 Charles Babbage Road, Cambridge CB3 0FS, UK

a)**Author to whom correspondence should be addressed:** aubin.prot@uni.lu

## ABSTRACT

Record efficiency in chalcopyrite-based solar cells Cu(In,Ga)(S,Se)₂ is achieved using a gallium gradient to increase the band gap of the absorber towards the back side. Although this structure has successfully reduced recombination at the back contact, we demonstrate that in industrial absorbers grown in the pilot line of Avancis, the back part is a source of non-radiative recombination. Depth-resolved photoluminescence (PL) measurements reveal two main radiative recombination paths at 1.04 eV and 1.5-1.6 eV, attributed to two phases of low and high band gap material, respectively. Instead of a continuous change in the band gap throughout the thickness of the absorber, we propose a model where discrete band gap phases interlace, creating an apparent gradient. Cathodoluminescence and Raman scattering spectroscopy confirm this result.

Additionally, deep defects associated to the high gap phase reduce the absorber performance. Etching away the back part of the absorber leads to an increase of one order of magnitude in the PL intensity, i.e., 60 meV in quasi Fermi level splitting. Non-radiative voltage losses correlate linearly with the relative contribution of the high energy PL peak, suggesting that reducing the high gap phase could increase the open circuit voltage by up to 180 mV.

## I. INTRODUCTION

Thin film solar modules based on Cu(In,Ga)(S,Se)$_2$ (CIGSSe) absorbers are already used in large scale production; AVANCIS/CNBM has a nominal production capacity of more than 1GW. They reach a record power conversion efficiency (PCE) of 20.3% for 30 cm × 30 cm submodules [1] and of 23.6% for cells. [2] The previous cell record of 23.4% [3] as well as the current submodule record make use of a double graded band profile. The front surface gradient is achieved by increasing the sulfur content at the front while the back surface gradient arises from a higher gallium content at the back of the absorber. Variation of the Ga/(Ga+In) ratio (GGI) over the thickness of the absorber has been studied for years. The resulting increasing band gap gradient towards the back side has been proven to help drive the carriers away from the back contact, thus preventing them from recombining non-radiatively before collection.[4–7]

Understanding the recombination activity in solar cells is essential, as non-radiative recombinations reduce the efficiency of the device. Carriers generated in the active material can either recombine radiatively, by emitting a photon with the energy of the transition, or non-radiatively, losing the energy to the lattice. In their famous publication on efficiency limits [8], Shockley and Queisser already addressed the impact of the non-radiative recombinations, demonstrating their detrimental effect on cell performance. A measure of the radiative recombinations and, by extension, of the non-radiative processes, is given by the photoluminescence quantum yield (PLQY, $f_c$ in the Shockley-Queisser paper), i.e., the ratio of the emitted photon flux over the incoming photon flux in a photoluminescence (PL) experiment. PL spectroscopy is a powerful and well known tool that can be used at an early stage in the preparation of solar cells to obtain valuable information about the quality of the absorber layer itself.[9] The current highest PLQY of a solar cell is obtained for GaAs cells, reaching 35.7%, whereas chalcopyrite based solar cells typically perform in the range 0.1-1%.[10]

All the absorbers investigated in this study are grown in the pilot line of AVANCIS, Germany. [11,12] They present an intentional double band gap gradient towards the front and back sides, induced by a sulfur and a gallium gradient, respectively. In this work, we compare the recombination activity from the front and back sides of these industrial Cu(In,Ga)(S,Se)$_2$ absorbers. We show by PL investigation that in graded absorbers, although the minimum of the band gap, i.e., the notch, dominates the recombination [13], there is an additional radiative recombination channel visible from the back of the absorbers. We propose here that the GGI does not gradually increase towards the back side as suggested by compositional profiles from Glow Discharge Optical Emission Spectroscopy (GDOES) measurements, but rather that two distinct phases coexist in the bulk of the absorber. A high density of high (low) GGI grains forms at the back (front) of the absorber and becomes sparser at intermediate depth where the two phases interlace. This leads to an apparent GGI gradient from a compositional point of view. We show further evidence supporting this hypothesis by cathodoluminescence spectroscopy and Raman spectroscopy.

## II.    METHODS

### Sample preparation

All the absorbers investigated in this study are based on a double barrier back-electrode for the control of alkali-diffusion and Mo selenization. They are grown in the pilot line of AVANCIS, Germany, according to the SEL-RTP process, i.e., stacked elemental layer followed by a rapid thermal processing. [11] The stacked elemental layer consists of precursor of Cu-In-Ga:Na deposited by sputtering and Se thermally evaporated. During the rapid thermal process in sulfur containing atmosphere, this layer reacts to form the CIGSSe absorber. The samples used in this study (with relative comparison summarized in Fig. 1b)) differ mostly for the In/Ga relative content during the precursor deposition, which consequently changes the overall GGI of the final absorbers, hence also the GGI at their back.

### Lift-off procedure

To gain direct access to the back side of the absorbers, a lift-off procedure is performed, removing the CIGSSe film from its molybdenum back contact, see Fig. 1. A two-component epoxy glue (Loctite EA 3421) is spread on the front surface of the sample before a clean glass is pressed on it. After at least 24h of curing, the samples are mechanically removed from their substrate by applying some force on them. We found that a short dip of the stack in liquid nitrogen (5-10 sec) beforehand helps lifting off the film more easily. It has been reported in the literature that the cleavage of such a stack happens at the interface CIGSSe/Mo or within the $Mo(S_x,Se_y)$ layer that forms at the interface. [14–16] X-ray diffraction (XRD) measurement on the remaining substrate – after lift-off – indicates that some residual traces of CIGSSe phase are present, suggesting that the current procedure does not provide an excellent detachment of CIGSSe from the Mo layer.

## Bromine etching

In order to investigate the absorbers via photoluminescence at different depths, we developed a bromine etching routine to etch part of the absorbers prior to the PL measurement. An aqueous bromine solution of concentration 0.1 M is prepared by diluting liquid bromine ($Br_2$) (0.3 mL) into distilled water (60 mL) and finally adding potassium bromide (KBr). The absorbers are dipped into the etching solution for 40s to 4 min depending on the desired etching depth. The remaining thickness of the etched absorbers is measured by cross-sectional SEM imaging. The etching is either performed on absorbers before the lift-off, to remove the front part of the absorber, i.e. the low GGI area, or performed after the lift-off, to remove the back part, i.e. the higher GGI area.

## Photoluminescence spectroscopy

We perform absolute calibrated PL measurements, exciting the samples with a 660 nm wavelength diode laser. Achieving the calibration is done in two steps: spectral correction of the measurements and calibration of the intensity. To fulfil the first point, a commercial halogen lamp of known emission (Avantes AvaLight-HAL-CAL-Mini) is directed onto a spectralon placed at the sample position and its spectrum is measured. Comparing the measurement and the calibrated spectrum provided by the company leads to a correction function that takes into account the spectral dependency of the collection system. The second point of the calibration requires on one hand to measure the beam radius and laser power at the sample position and compute the resulting expected flux, and on the other hand to measure its spectrum via the collection system. Comparing the measured (and spectrally corrected) flux with the expected one returns an intensity correction factor. If not specified otherwise in the text, the laser output power is set in such a way that the impinging photon flux is $2.9 \cdot 10^{17}$ photons/cm²/s, corresponding to a one sun generation for a material of band gap 1.04 eV, based on the AM1.5 solar spectrum. The emission of the sample is directed into an Andor Shamrock sr-303i spectrograph via two parabolic mirrors that focus the emitted photons into an optical fiber (0.22NA, 550 μm diameter) and is finally detected by a CCD Si camera (Andor iDus DV420A-OE) in the visible range and a CCD InGaAs camera (Andor iDus DU490A-1.7) in the NIR range. For low temperature measurements, the samples are placed in an Oxford Instruments cryostat (OptistatCF) and cooled down to temperatures as low as 10 K using liquid helium as cryogenic agent. Some samples require to be measured through glass. However, excited with a 660 nm laser, glass emits radiation at 1.5 eV which adds unwanted contribution to the PL spectra. Therefore, a preliminary measurement of the glass emission alone is required and can be subtracted from the absorber spectrum (see Fig. S1 in SM).

## Glow discharge optical emission spectroscopy

Depth profile analysis by Glow Discharge Optical Emission Spectroscopy (GDOES) was applied to evaluate the chemical composition within the CIGSSe absorbers. The system uses an Ar-plasma for sputtering and a CCD-array to detect the photons emitted during the relaxation of the sputtered and in the plasma excited atoms and ions. All GDOES depth profiles were measured using a GDA 650 HR, a system built by Spectruma Analytik in DC excitation mode (constant voltage-constant current mode) including an anode with an inner diameter of 2.5 mm. The excitation parameters were set to 1000 V and 12 mA. The WinGDOES software was applied to automatically determine quantitative depth profiles of mass concentration and then correct the data with a calibration method developed internally.

## Raman spectroscopy

Raman spectroscopy measurements are performed in a backscattering configuration through a probe designed at IREC using 442, 532 and 785 nm excitation wavelengths. For the 532 and 785 nm excitation wavelengths, the measurement spot diameter is ~70 μm (macro-Raman) and up to 25 points were measured over all the surface of both front and back sides, which provided representative information for the whole sample. For the 442 nm excitation wavelength, the measurement spot diameter is ~2 μm (micro-Raman) and 64 points are measured on a 3×3 mm$^2$ area from the back side. This allows to evaluate the inhomogeneities at the micro-scale and acquire information from individual grains. For the 442 and 532 nm excitation wavelengths, a fHR 640 monochromator from Horiba Jobin Yvon, coupled with a CCD detector cooled down to -130˚C, is used. For the 785 nm excitation wavelength, an iHR 320 monochromator from Horiba Jobin Yvon, coupled with a CCD detector cooled down to -70˚C, is used. To avoid the presence of thermal effects in the spectra, excitation power density is kept below 50 W·cm$^{-2}$ in the laser spot. All spectra are calibrated by measuring a reference monocrystalline silicon sample and imposing its main Raman peak to 520 cm$^{-1}$.

**Cathodoluminescence spectroscopy**

CL hyperspectral mapping is performed at 300 K, on cleaved cross-section in an Attolight Allalin 4027 Chronos SEM-CL system. CL measurements are taken using an iHR320 spectrometer with a grating density of 150 lines per mm blazed at 500 nm. The microscope is operated at an electron beam current of 10 nA and an acceleration voltage of 8 kV. The CL hyperspectral maps are then analyzed using LumiSpy. [17]

## III.   RESULTS AND DISCUSSION

The composition profile of the studied absorbers is obtained by means of GDOES, from which band gap profiles can be computed according to Bär *et al.*.[18] Typical band gap and GGI profiles of the absorbers from this study are plotted in Fig. 1a). However, we empirically observe that the minimum of the band gap energy ($E_{g,min}$) given by the GDOES profile systematically leads to higher values compared to other measurements methods. Comparing $E_{g,min}$ of the studied absorbers (see Fig. S2 in SM) obtained from reflection, PL, EQE  and GDOES analysis [9,18–21] reveals that the first three methods agree on $E_{g,min}$, whereas GDOES overestimates it by about 55 meV. We believe the reason for this is that not all of the sulfur is incorporated in the grains, and that non-negligible amounts of it are located at the grain boundaries. Keller *et al.* investigate the atomic structure around grain boundaries in AVANCIS' CIGSSe absorbers by Atom Probe Tomography (APT) and shows that sulfur tends to agglomerate at the grain boundaries. [22] Consequently, band gap energies calculated from composition alone are overestimated due to the apparent higher sulfur content. Additionally, as it will be discussed in further detail later, Raman spectroscopy reveals some sulfur-free areas on the back side. Therefore, we will only consider the band gap profiles in a qualitative way or refer directly to the GGI profiles.

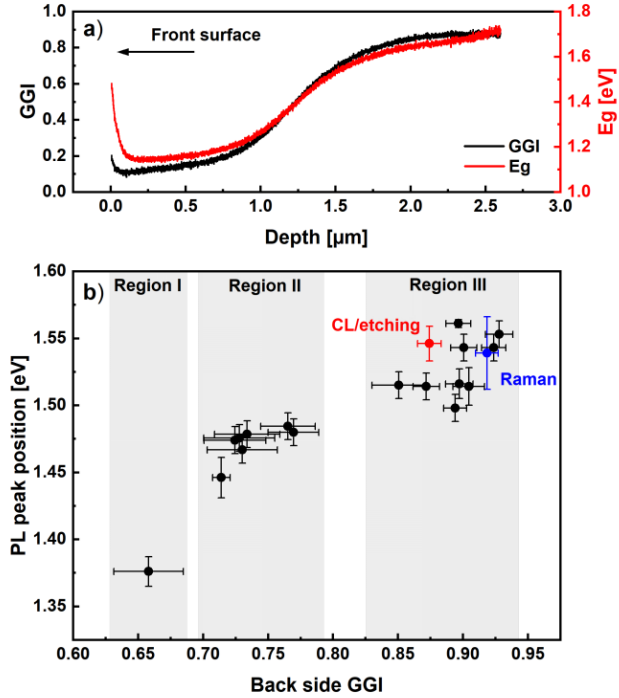

FIG. 1. a) Typical GGI (black) and band gap (red) profiles of an absorber from region III (red dot) determined from GDOES compositional measurements. The investigated absorbers are all about 2.6 μm thick. The stronger increase of the band gap from GDOES towards the front is due to the sulfur content, which is higher near the front. b) 19 samples are investigated. For each of them, the position of the high energy peak is measured by PL from the back side. The error in PL energy is due to inhomogeneities and is determined from different measurements on the same sample. The back side GGI is determined from an average over the last 400 nm of the corresponding GDOES GGI profile. The standard deviation is taken as the error. The two highlighted samples are discussed later in the text.

## A. A high energy peak in the photoluminescence spectra

In order to investigate the recombination activity from the back side of the different samples, the absorber is mechanically removed from its substrate, as described in the methods section, and a subsequent analysis of the PL response from the different faces is conducted. PL spectra measured from the back side show two main emission peaks: one peak centered around 1.03 eV and another peak in the range of 1.5-1.6 eV (see blue spectrum in Fig. 2). The small sharp peak at 0.89 eV is attributed to water absorption [23] and there is a second broader contribution centered around 0.9 eV that will be discussed in further detail later.

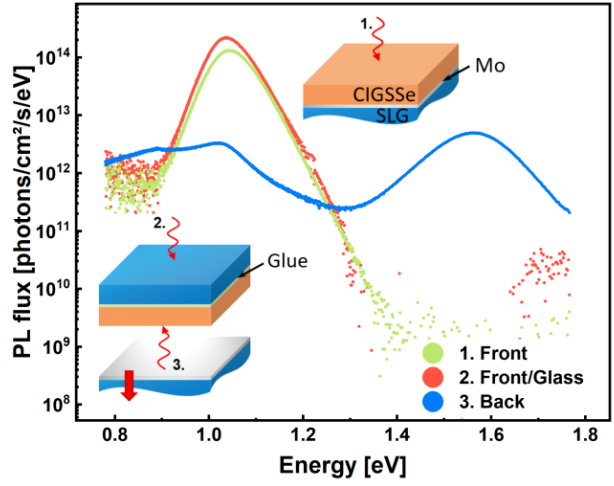

FIG. 2. Semi-log scale – absolute photoluminescence spectroscopy at the different stages of the lift-off process: under front side illumination before (1.) and after (2.) lift-off, respectively and under back side illumination after lift-off (3.). Spectrum 2 has been corrected from the glass emission as indicated in the methods section.

At room temperature, however, measurements with illumination from the front side, before or after the lift-off (red and green spectra in Fig. 2), reveal only one single contribution at about 1.04 eV. This PL emission peak corresponds to band-to-band recombination in the region of the band gap minimum of the absorber. [9,13] We therefore argue that from both front and back side illumination, significant recombination originates from the band gap notch. After the lift-off, the stress applied on the lattice by the molybdenum is released, allowing the lattice to relax. Thus, we generally observe that the notch PL peak is 10-20 meV lower after lift-off than before, depending on the curing time. [13] The longer the glue cures before the lift-off, the less shift is observed. Moreover, as shown in Fig. 2, the notch luminescence is about two orders of magnitude lower when excited and measured from the back side than from the front side as a consequence of the second recombination channel. According to the band gap profile shown in Fig. 1, one would expect a higher band-to-band emission while measuring from the back side. However, this model cannot explain why only two peaks are detected. If one assumes that the high energy peak originates from the band gap gradient at the back side, then all the intermediate band gaps should also be visible by PL and result in one single very broad peak. Unexpectedly, the high energy peak can be detected from the front side as well, but only at low temperatures (in the range of 10-80 K, see Fig. S3 in SM). At low temperatures, a decrease of the photogenerated carriers' mobility is expected and therefore the recombination is more likely to occur at the generation site. Moreover, the absorbers are excited with a laser of wavelength 660 nm having a penetration depth in the material of about 100 nm. It suggests that within the first few hundreds of nanometers from the front surface, the second recombination channel already exists, emitting at 1.5-1.6 eV.

In order to understand the origin of the high energy peak, we prepared a series of absorbers with varying GGI at the back side. As discussed above, we believe that the sulfur is not entirely incorporated in the grains and therefore we use the back side GGI as a guide to compare the band gap energy trends between the samples. Note that this does not provide an exact value for the expected band gaps. Fig. 1b) shows the relation between the position of the high energy PL peak measured from the back side and the back side GGI averaged over the last 400 nm of the GDOES profiles. Samples in region III present small changes in back side GGI, ranging from 0.85 to 0.93 and only a loose correlation is observed for the corresponding high energy PL peak, which varies from 1.50 eV to 1.56 eV. Samples in region II have a significantly reduced back side GGI of 0.71 to 0.77. The same PL peak energy is observed within errors, except for one sample with slightly lower peak energy (and lower back side GGI). Lastly, the sample in region I has the lowest back side GGI (0.66) and shows the lowest peak energy (1.38 eV). This demonstrates that the band gap of the high gap phase strongly depends on the GGI at the back side of the absorber.

Based on these results, we argue that there is no continuum of band gaps through the material, as suggested by the GDOES profile in Fig. 1a), but rather that there are distinct phases of discrete band gaps coexisting throughout the whole thickness of the material. Moreover, the position of the high energy PL peak depends on the gallium content at the very back of the absorbers.

In a previous work on co-evaporated CIGSe absorbers from ZSW, Wolter *et al.* detected a similar high energy peak in addition to the notch peak when measuring from the back side. [13] These samples were grown with a back side GGI of about 0.6 and the high energy peak was detected at 1.31 eV, correlating with the trend observed in this work.

## B. Distinct phases of discrete band gaps

Investigation by Raman spectroscopy of both the front and back sides of the absorbers is conducted to assess whether the different phases are due to different CIGSSe compositions or due to secondary phases. In addition, cross sectional cathodoluminescence provides local information about the band gap throughout the film thickness.

Raman spectroscopy measurements are performed under a 532 nm laser excitation in a macro configuration (see methods section for details). This excitation wavelength is close to the band gap of pure $CuGaS_2$ [24], which leads to resonant Raman scattering and a strong increase of the LO components of the polar E/B symmetry modes. [25] Although four absorbers were measured (all from region III in Fig. 1b)), Fig. 3a) only shows the normalized Raman spectrum of one of them, as no significant differences could be observed between the samples (blue dot in Fig. 1b)). The main peaks, observed at the front side, are associated to the A-like symmetry modes, i.e., Se-Se and S-S vibrations at 178 cm$^{-1}$ and 288 cm$^{-1}$, respectively. [26,27] When measuring from the back side, a slight blueshift of these peaks to 185cm$^{-1}$ and 307 cm$^{-1}$, respectively, is observed (see Fig. 3b)). This can be explained by the increase in sulfur and gallium content towards the back of the absorbers. [26,28] The associated increase in the band gap also leads to resonant behavior of the spectra measured from the back side, with a high relative intensity of the LO components of the E/B-like symmetry modes (indicated by the green dashed lines in Fig. 3a) and b)). The peaks marked by the gray dashed lines, at 170 cm$^{-1}$, 238 cm$^{-1}$, and 252 cm$^{-1}$, are attributed to a $MoSe_2$ phase detected at the back side of the absorbers. [29] This is an expected secondary phase at the Mo-CIGSSe interface. All other peaks are due to CIGSSe phases, no additional secondary phases are observed. An investigation of the remaining substrate after lift-off reveals some traces of CIGSSe phases, indicating the imperfect removal of the absorber from its substrate.

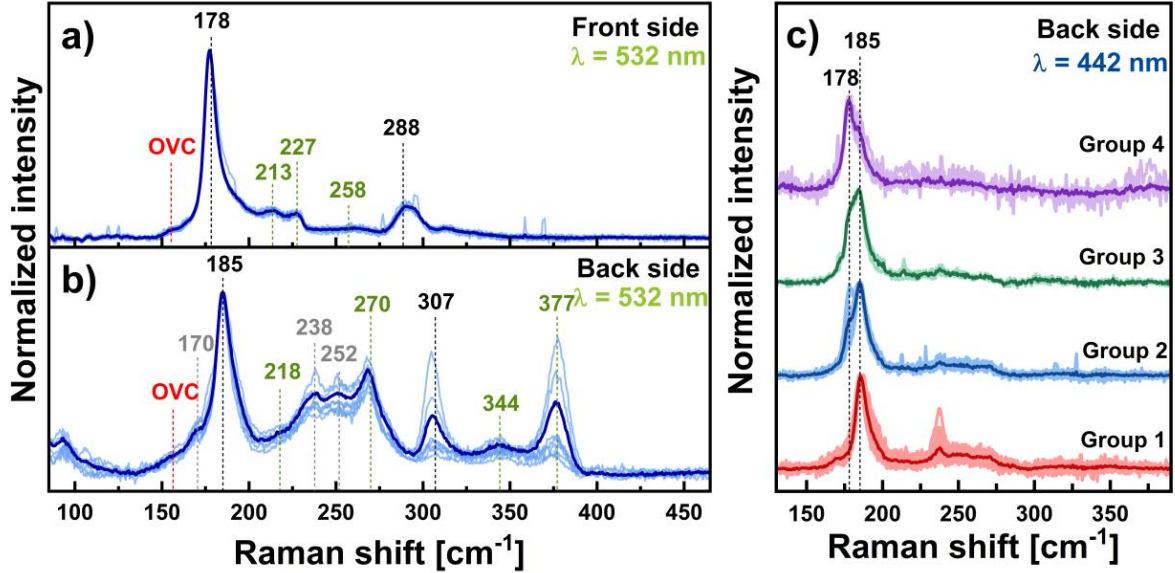

FIG. 3. Typical Raman spectra measured from a) the front side before lift-off and b) the back side after lift-off. A 532 nm wavelength laser is used as an excitation source. c) Micro Raman on a 3×3 mm$^2$ area from the back side using a 442 nm wavelength laser as excitation source. Individual measurements are ordered in group 1 to 4 depending on the dominant peak. Pale lines in all figure parts are individual measurements taken on different positions and normalized to the peak related to Se-Se vibrations (178 cm$^{-1}$ and 185 cm$^{-1}$, respectively). Darker lines are the resulting averaged spectra.

Additionally, a small peak of an ordered vacancy compound (OVC) phase is visible from both sides, appearing as a shoulder on the low frequency side of the Se-Se peak, as highlighted in Fig. 3a) and b) by the red dashed lines. [30] Under excitation of a 785 nm laser, more of the OVC phase peaks are detected (see Fig. S4 in SM) from both front and back sides due to this wavelength being close to resonance for the CuIn$_3$Se$_5$ OVC phase. [31] In Cu-poor chalcopyrites, the formation of off stoichiometry compounds, i.e., OVCs, is often observed, such as CuIn$_3$Se$_5$ or CuIn$_5$Se$_8$ phases for instance, in selenide absorbers. [32,33]

Furthermore, the pale lines in Fig. 3a) and b) represent individual measurements performed at different positions on the absorber. They can be used as an indication of the lateral homogeneity, in terms of chemical composition, of the absorbers' surfaces. It follows that absorbers are very homogeneous from the front side, whereas significant variation is observed between the individual measurements from the back side, suggesting larger inhomogeneities. In particular, as it can be concluded from the absence of the 307 cm$^{-1}$ peak in Fig. 3c), some sulfur-free areas are present on the back side. More interestingly, grains with significantly different gallium amount can be detected. The back side of the absorber studied in Fig. 3a) and b) is additionally scanned over an area of 3×3 mm$^2$ with a 442 nm wavelength laser in micro configuration (see methods section), which enables a grain to grain analysis (Fig. 3c)). This excitation wavelength has been chosen for being far from resonant with pure CuGaS$_2$, CuGaSe$_2$ and CuIn$_3$Se$_5$ OVC phases [24,31,34] and has a penetration depth shorter than 100 nm. Two distinct Cu(In,Ga)Se$_2$ phases could be distinguished when measuring from the back side, one more Ga-rich and the other more In-rich, yielding the peaks at 185 cm$^{-1}$ and 178 cm$^{-1}$, respectively. For clarity, the individual measurements are divided into four groups depending on the dominating phase. The Ga-rich phase strongly dominates in groups 1 and 2, whereas the In-rich dominates the spectra in group 4. Group 3 regroups spectra where the two phases compete.

Complementary information on the variation of the band gap from grain to grain can be obtained from cathodoluminescence spectroscopy (CL) in the scanning electron microscope (SEM). In particular, CL is performed on the cross section of an absorber similar to the one in Fig. 3 (red dot in Fig. 1b)). In each pixel, a spectrum is measured making it possible to follow the evolution of the band gap from the front surface down to the back side. Fig. 4a) shows a secondary electron (SE) cross section image of the region where CL is performed and Fig. 4b) reports the energy of the maximum of each individual spectrum. It is important to mention that this representation gives only a simple view of the band gap energies, as some spectra, especially close to the back side, are too noisy to identify peaks. As the sample is slightly tilted, not only the cross section's emission is visible, but also the very front surface. The latter is dominated by a low gap phase with emission at about 1.13 eV (blue color). We relate this low gap phase to the low energy peak in the PL spectra, centered at 1.04 eV. Energy differences between the two measurement methods is attributed to the decreased quantum efficiency of the Si-camera used in CL at such low energies.

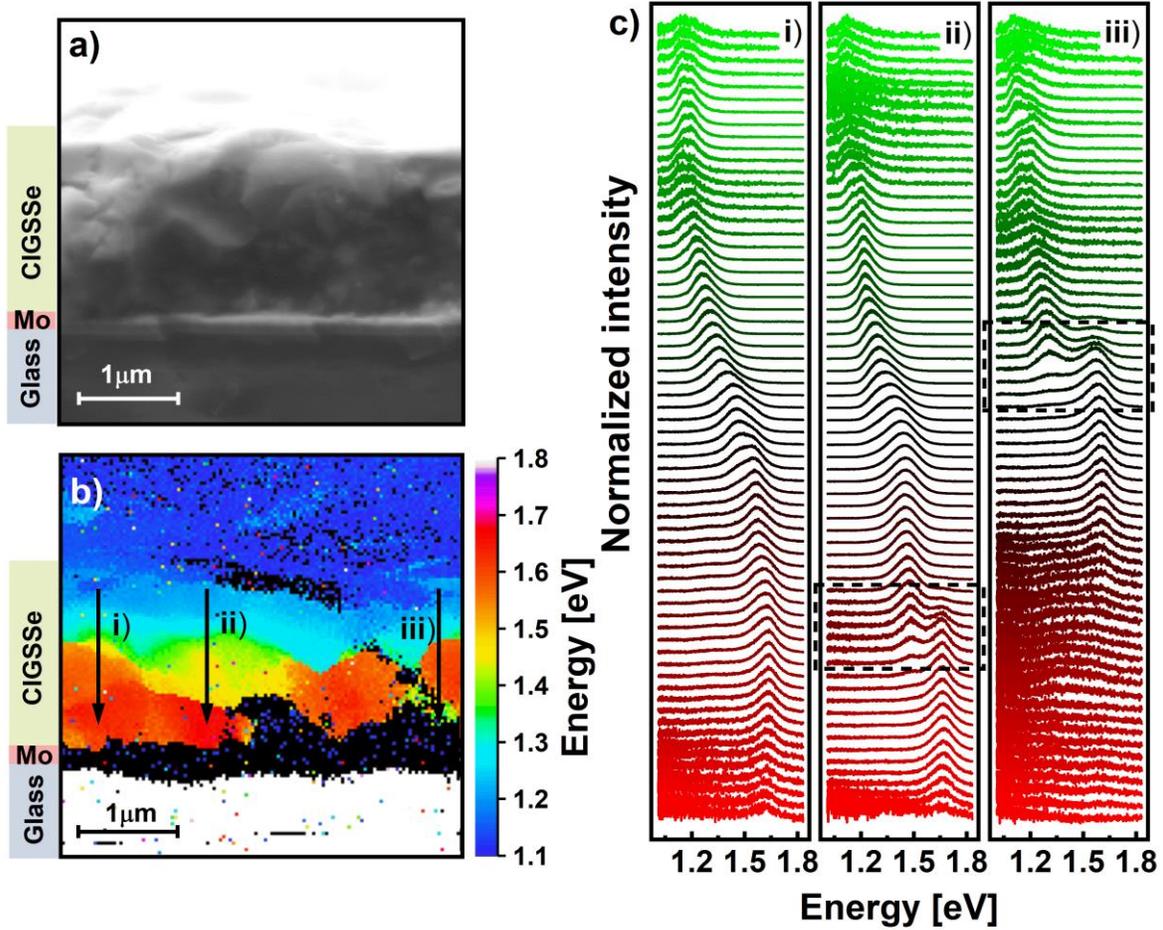

FIG. 4. a) SE cross section image of the region of interest where the CL is performed. b) Colormap of the energy of the maximum of the CL emission. Orange-red corresponds to the high gap phase, green-yellow to the intermediate one and blue-cyan to the low gap phase. The white region at the bottom is glass and the black regions are either molybdenum (see left legend) or too noisy to identify peaks. c) individual spectra along the lines i-iii) shown in b). The peak in the transition region broadens compared to the region where only one peak is observed. The dashed boxes indicate the coexistence regions.

Considering individual line profiles from top to bottom along the cross section of the film, we identify three scenarios, as indicated in Fig. 4b) by the vertical lines: i) two phase region with an almost smooth transition from the front surface towards the back side, ii) three phase region with an almost smooth transition at intermediate depth of low gap and intermediate gap phase and with coexistence deeper of intermediate and high gap phase, iii) two phase region with coexistence in the middle of the film between low gap and high gap phase. Individual spectra along the three lines are plotted in Fig. 4c). In case i) only one emission peak is detected, gradually increasing from 1.15 eV to 1.62 eV towards the back side. It should be noted, however, that in the transition region the peak becomes wider than in the regions where clearly only one phase is detected. Cases ii) and iii) show distinct areas, indicating multiple phases. In case ii), a gradual increase similar to the one from i) is observed from 1.13 eV until 1.48 eV, but a sharp transition from 1.48 eV to 1.67 eV then occurs, as can be seen from the intermediate spectra where two peaks occur in the spectra. The low energy peak loses in intensity as the higher energy one takes over. This abrupt transition from one recombination channel to another is observed in the colormap (Fig. 4b)) as an abrupt change of color. The line profile iii) is another example of such an abrupt change of recombination energy, happening however closer to the front surface, changing the dominant recombination from 1.28 eV to 1.61 eV. From a larger set of CL measurements (see Fig. S5 in SM), it is seen that scenario i) is not the most probable, but that generally, one or two abrupt transitions occur, indicating two or three discrete phases.

### C. Depth resolved investigation of the distinct interlaced phases

To support the hypothesis that the bulk of the absorber is made of interlaced distinct phases, depth resolved photoluminescence spectroscopy is performed, after sequential etching of the absorber. Using the bromine etching routine described in the methods section, we gradually etch an absorber from region III (red dot in Fig. 1b)) and measure the photoluminescence response at different depths throughout the thickness. Thus, information about the depth distribution of the different phases can be accessed. Etching from the front and back side has been performed and the remaining thickness of the absorbed is measured via SEM cross section. The etched surface generally presents an unequal roughness and therefore we estimate the error in the measurement of the thickness to be ±0.2 μm.

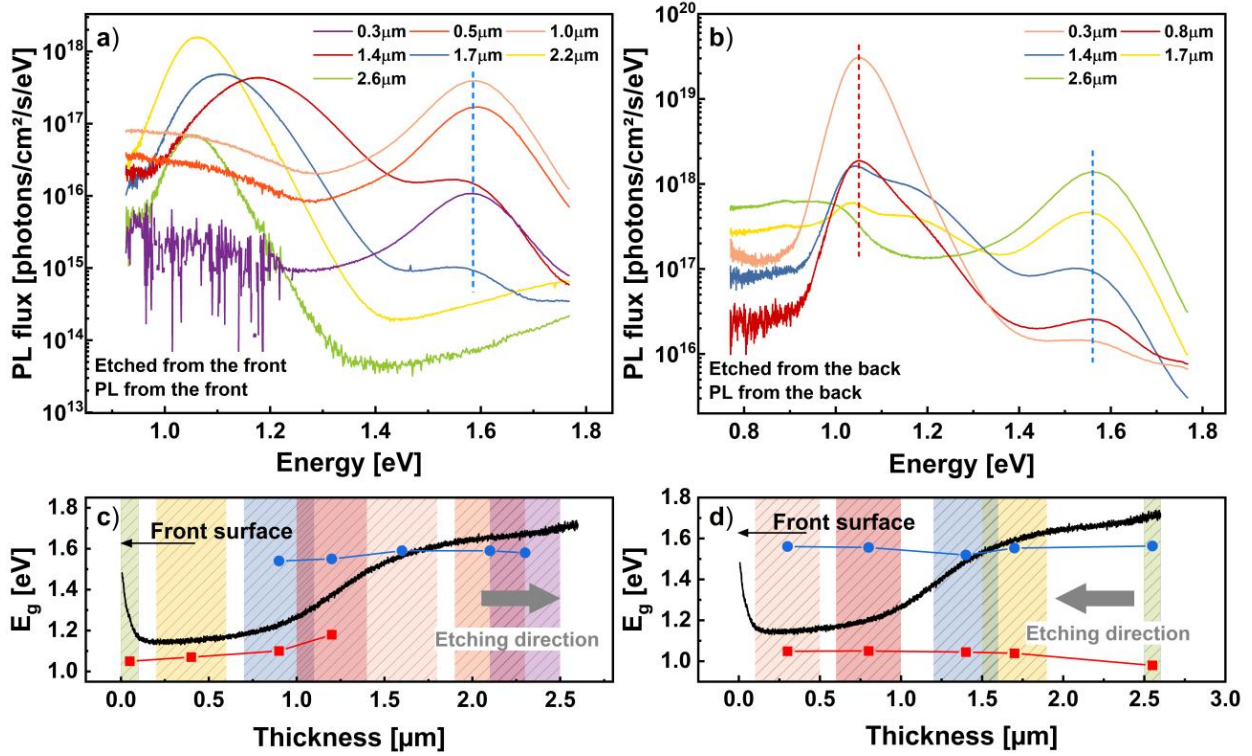

FIG. 5. Top: non-absolute photoluminescence spectra at different depths after the absorber is gradually etched, from the front side a) and from the back side b). The legend indicates the remaining thickness after the etching and the dashed lines represent the peaks' position, also reported in c) and d). Bottom: band gap profile determined from GDOES (black) and energies of the measured PL notch peak (red) and high energy peak (blue) at the different etching steps, from front side c) and back side d). The colored bars indicate the etching depth, including errors.

After the first etching step from the front side, about 400 nm of material is etched and the notch peak intensity increases by one order of magnitude. The low intensity of the non-etched absorber is attributed to the degraded surface after air exposure, which is a region of non-radiative recombinations. [35,36] We empirically notice that a buffered absorber remains stable for months, whereas a bare one oxidizes in a few hours in the air, resulting in a decrease of the PL intensity by up to two orders of magnitude. It should also be noted that even though there is a steep $E_g$ increase at the very front surface of the absorbers, the expected recombination, when no etching is performed, is still at the band gap minimum, due to carriers' diffusion. As more of the material is removed, we observe that the peak related to the notch recombination reduces in intensity and blueshifts slightly before disappearing entirely (see Fig. 5a)). This behavior correlates well with the band gap profile shown in Fig. 5c) and confirms that the dominant recombination occurs at the notch. As previously discussed, the GDOES profile should be considered in a qualitative way to address trends and not as offering absolute values, as we believe that not all the sulfur is in the bulk of the grains. On the other hand, before the notch peak disappears, the high energy peak, centered at 1.56 eV, is detected after 0.9μm of the absorber is etched (remaining thickness 1.7 μm) and remains at constant energy until only 300 nm of material are left. This is in contradiction to the smooth profile indicated by GDOES and in agreement with the cathodoluminescence observations which have already shown that some high band gap material is present in the middle of the absorber and not solely at the back side.

To gain a better understanding of which contribution comes from what depth, we perform the etching (as well as the PL measurement) from the back side of the absorbers, thus conserving the notch at all times and removing only the back part (see Fig. 5b) and d)). Therefore, no shift of the notch emission peak is observed, as expected. In a similar manner as from the front side, as more of the material is etched away, the high energy peak remains at constant energy. Additionally, it gradually loses in intensity, while the notch peak's intensity increases, confirming that the high gap phase is more present towards the back side of the absorber. Moreover, an additional PL peak at about 1.2 eV is detected at intermediate depth, in the absorber's bulk (at remaining thicknesses 1.7 μm and 1.4 μm) from the back side, as it can be seen in Fig. 5b). We attribute this peak to a third phase of intermediate $E_g$, as already seen in CL spectroscopy (Fig. 4 line profile ii)). In addition, the broadening of the notch peak (at remaining thicknesses 1.7 μm and 1.4 μm), when measuring from the front side, is another indication of the presence of this phase. While cathodoluminescence provides information on the different phases present throughout the absorbers on a micrometer scale, photoluminescence collects signal from a few hundred grains at once. We therefore argue that depth resolved PL spectroscopy gives an average of the three scenarios discussed in Fig. 4, showing that mainly two or three individual phases with different GGI ratio are formed within the material. We propose in Fig. 6 a new model for the band gap profile in such graded absorbers. According to GDOES measurements, a gradual increase from a low band gap energy (green in Fig. 6a)) towards a high band gap energy (light red) is expected. However, as we demonstrated in this work, a situation as depicted in Fig. 6b) seems more reasonable. A large density of high gap phase, the $E_g$ of which is influenced by the GGI at the back side, forms close to the back contact of the absorber. Contrastingly, a low gap phase forms in high density towards the front side, generating the "notch" part of the absorber. In the bulk of the material, these two phases coexist, in different densities, giving rise to an apparent gradient in the GDOES profile. In addition to the two main phases, a sparse phase of intermediate band gap could be detected in the bulk, represented by the dark red color in Fig. 6b). Finally, as observed by CL, some rare grains show a smooth $E_g$ gradient as highlighted on Fig. 6b).

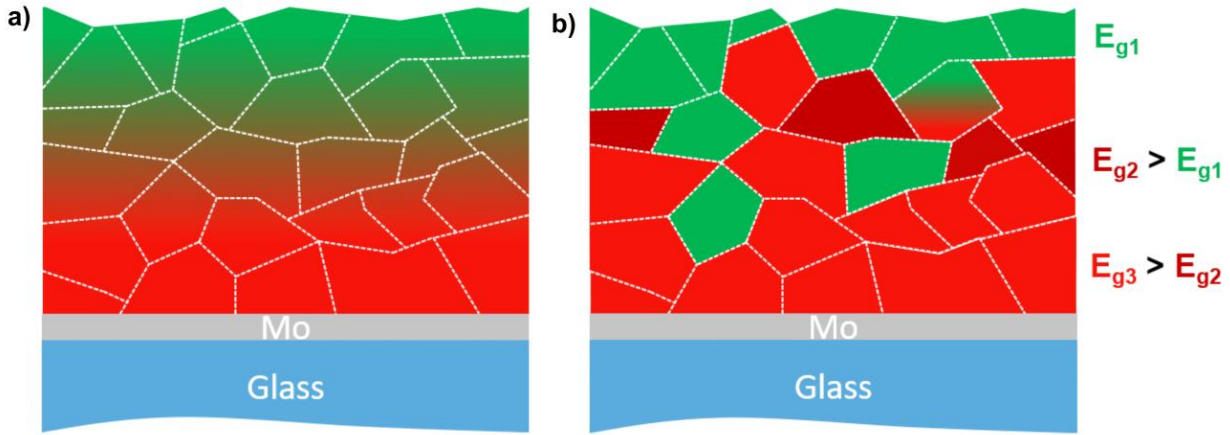

FIG. 6. Model for gallium gradient as suggested from GDOES profile a), and as proposed in this work based on other techniques b). Green indicates a low band gap phase ($E_{g1}$), dark red an intermediate band gap phase ($E_{g2}$) and light red a high band gap phase ($E_{g3}$).

## D. Effect of the back side luminescence on the radiative efficiency

Interestingly, as we etch from the back side past the steep Ga gradient and most of the high gap phase is removed (last etching step from the back side, with remaining thickness 0.3 μm), the notch peak's intensity increases by one order of magnitude. Its total integrated flux is $3 \cdot 10^{18}$ photons/cm²/s. All other spectra in Fig. 5b) have similar total integrated flux of about $3 \cdot 10^{17}$ photons/cm²/s. This strongly indicates that the non-radiative losses decrease when the back part is removed, i.e., that the back part of the absorber provides undesirable recombination paths that are detrimental to the absorber. A similar observation has been made based on carrier lifetime in graded CIGSe absorbers. [37] In fact, an increase of one order of magnitude in PL intensity translates into an increase of approximately 60 meV in quasi Fermi level splitting, which gives an upper limit for the $V_{OC}$ of the complete cell. [9] Among the undesirable recombination paths, a broad PL contribution centered at about 0.9 eV is detectable from the back side, and from the front side after etching (see Fig. 2 and 5). Other chalcopyrite thin films studies reported on a PL emission at such low energy. Niki *et al.* observed a broad peak at about 0.85 eV in Cu-poor and Cu-rich CuInSe$_2$ films. [38] Mansfield *et al.* concluded that in standard Cu-poor Cu(In,Ga)Se$_2$ a band-to-defect emission was detectable by PL spectroscopy at 0.79 eV. [39] More recently, Siebentritt *et al.* showed that similar deep defect is also detected in sulfide chalcopyrites. [40] Spindler *et al.* reported on a defect transition in Cu(In,Ga)Se$_2$ around 0.7 eV to 0.8 eV, for high GGI and low GGI, respectively. [41] However, similarly to the observations in [38], they showed that the transition was more pronounced in Cu-rich films than in Cu-poor samples. They also discussed deep broad transitions around 1.10 eV and 1.24 eV in CuGaSe$_2$ at 10 K that they tentatively attributed to Ga$_{Cu}$ and [Ga$_{Cu}$-2V$_{Cu}$], respectively. [42] In the present work, the 0.9 eV defect emission may be a combination of the 0.8 eV and 1.10 eV defects. Furthermore, this defect emission disappears as the back part of the absorber is etched away, hinting that these defects are associated to the high gap phase, present in higher density at the back side. This defect disappearing could be the cause of the increase in PL intensity observed in Fig. 6b) after the last etching step. Two additional absorbers (from region III, Fig. 1b)) have been gradually etched from the back as well, but with shorter etching steps past the steep Ga gradient (see Fig. S6 in SM). It shows more clearly that the low energy PL peak intensity gradually increases as the back side is etched away. However, we cannot exclude the contribution of the interface between the low and high band gap phases. When etching enough of the back side, less high band gap material is left in the remaining absorber, reducing therefore the number interfaces between low and high band gap phases. This could in turn increase the PL intensity of the low energy peak.

To investigate the influence of the high gap phase on the radiative efficiency of the absorbers, we consider different CIGSSe absorbers and measure for all of them absolute photoluminescence from both front and back sides. On one hand, based on the measurements conducted on the front side, we extract the non-radiative losses as $k_b T \times \ln(\text{PLQY})$, where $k_b$ is the Boltzmann constant, $T$ the temperature of the measurement and PLQY the photoluminescence quantum yield. On the other hand, based on the measurements from the back side, we determine the relative contribution of the high energy peak to the notch peak. In other words, we calculated the ratio of the integrated fluxes of the high energy peak and of the notch peak, referred as "peaks ratio" in Fig. 7. A linear relation between the non-radiative losses and the peaks ratio is obtained, demonstrating the negative impact of the high gap phase on the device's performance. We show that high non-radiative losses are associated with high peaks ratio and conclude from the fit that decreasing the amount of the high gap phase could reduce these losses by up to 180 meV.

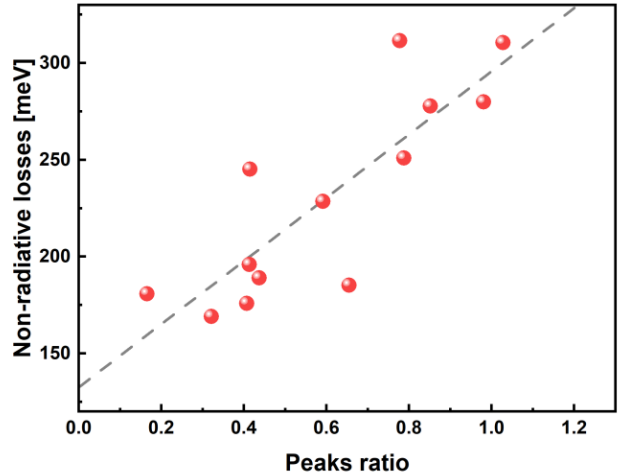

FIG. 7. Non-radiative losses vs. peaks ratio for different CIGSSe absorbers. The peaks ratio is calculated as the ratio of the integrated PL fluxes of the high energy peak and of the low energy peak, from measurements from the back side. The dashed line is a linear fit to the data.

## IV. CONCLUSION

We demonstrated that the back side of graded $Cu(In,Ga)(S,Se)_2$ absorbers is a source of non-radiative recombinations and therefore a limiting parameter for cells' performances. Depth resolved photoluminescence spectroscopy revealed that two main radiative recombination paths compete: one at about 1.04 eV, corresponding to recombination in the low gap phase and one around 1.56 eV, attributed to recombination in the high gap phase. At intermediate depth into the bulk of the absorber, a third recombination path can sometimes be detected, indicating the presence of a sparse phase of band gap around 1.2 eV. Raman and cathodoluminescence spectroscopy confirm the formation of multiple phases and locate the low (high) band gap phase close to the front (back) side. We propose that the band gap gradient, in graded absorbers, is not a gradual increase from the front side towards the back side, but rather that two main phases of low and high band gap form, with the low gap phase predominantly near the front and the high gap phase predominantly near the back. The interlacing of these two phases in the bulk of the material produces the apparent gradient one can measure with GDOES.

In addition, we detect a broad defect band centered around 0.9 eV inside the high gap phase. We tentatively propose, as an origin, a combination of deep defects, one around 0.8 eV, commonly observed in selenide and sulfide chalcopyrites, and another one around 1.1 eV, as detected in CGSe. By etching the back side of the absorber until most of the high gap phase is removed, we measured a strong increase of one order of magnitude in the PL intensity of the notch peak, which translates in an increase of 60 meV in quasi Fermi level splitting. Furthermore, we find a linear relation between the non-radiative voltage losses and the contribution of the high energy PL peak, highlighting the detrimental effect of the high gap phase. A decrease of the non-radiative losses of up to 180 meV could be expected by reducing the amount of the high gap phase.

## SUPPLEMENTARY MATERIAL

See the supplementary material for complementary PL, CL and Raman data as indicated in the text and details on band gap determination based on different measurement methods.

## ACKNOWLEDGMENTS

This work has been supported by Avancis, Germany, in the framework of the POLCA project, and by EPSRC in the framework of the REACH project (EP/V029231/1), which are gratefully acknowledged. Authors from IREC belong to the MNT-Solar Consolidated Research Group of the "Generalitat de Catalunya" (ref. 2021 SGR 01286) and are grateful to European Regional Development Funds (ERDF, FEDER Programa Competitivitat de Catalunya 2007–2013). M.G. acknowledges the financial support from Spanish Ministry of Science, Innovation and Universities within the Juan de la Cierva fellowship (IJC2018-038199-I).

## AUTHOR DECLARATIONS

### Conflict of interest

The authors declare no conflict of interest.

### Author contributions

**Aubin JC. M. Prot:** Conceptualization (equal); Investigation (lead); Methodology (equal); Data Curation (lead); Formal analysis (lead); Visualization (lead); Writing/Original Draft Preparation (lead); Writing/Review & Editing (equal). **Michele Melchiorre:** Investigation (supporting); Formal analysis (supporting). **Felix Dingwell:** Investigation (supporting); Writing/Review & Editing (supporting). **Anastasia Zelenina:** Data Curation (equal); Formal analysis (equal);

Investigation (equal); Methodology (equal); Resources (equal). **Hossam Elanzeery:** Data Curation (equal); Formal analysis (equal); Investigation (equal); Methodology (equal); Resources (equal); Writing/Review & Editing (equal). **Alberto Lomuscio**: Data Curation (equal); Formal analysis (equal); Investigation (equal); Methodology (equal); Resources (equal); Writing/Review & Editing (equal). **Thomas Dalibor**: Formal analysis (equal); Resources (equal); Writing/Review & Editing (equal); Funding Acquisition (lead). **Maxim Guc**: Data Curation (equal); Formal analysis (equal); Investigation (equal); Writing/Review & Editing (equal). **Robert Fonoll-Rubio**: Data Curation (equal); Formal analysis (equal); Investigation (equal); Writing/Review & Editing (equal). **Victor Izquierdo-Roca**: Formal analysis (equal). **Gunnar Kusch**: Data Curation (equal); Formal analysis (lead); Investigation (lead); Writing/Review & Editing (equal). **Rachel A. Oliver**: Conceptualization (lead); Formal analysis (equal); Writing/Review & Editing (equal); Funding Acquisition (lead). **Susanne Siebentritt**: Conceptualization (lead); Formal analysis (equal); Funding Acquisition (equal); Methodology (equal); Supervision (lead); Writing/Review & Editing (equal).

## DATA AVAILABILITY

The data that support the findings of this study are openly available in Zenodo at https://doi.org/10.5281/zenodo.8017507.

# Supplementary Material

## Composition variations in Cu(In,Ga)(S,Se)₂ solar cells: not a gradient, but an interlaced network of two phases

Aubin JC. M. Prot, Michele Melchiorre, Felix Dingwell, Anastasia Zelenina, Hossam Elanzeery, Alberto Lomuscio, Thomas Dalibor, Maxim Guc, Robert Fonoll-Rubio, Victor Izquierdo-Roca, Gunnar Kusch, Rachel A. Oliver, Susanne Siebentritt

### S1.   Photoluminescence measurement through glass

Measuring from the front side after performing the lift-off of the film means that the laser first goes through a slide of soda lime glass (SLG). Under excitation of a 660 nm laser, we have found that SLG emits at 1.5 eV, which distort greatly the PL spectrum of the absorbers (see blue "non-corrected" spectrum in Fig. S1). However, taking a PL measurement of the glass alone under the same conditions as the absorber's makes it possible to correct the shape (black dashed spectrum). A simple subtraction of the glass spectrum from the absorber spectrum is performed. The quality of the subtraction can be assessed from the slope of the high energy wing (0.1 eV above the band gap energy) of the red "corrected" spectrum.

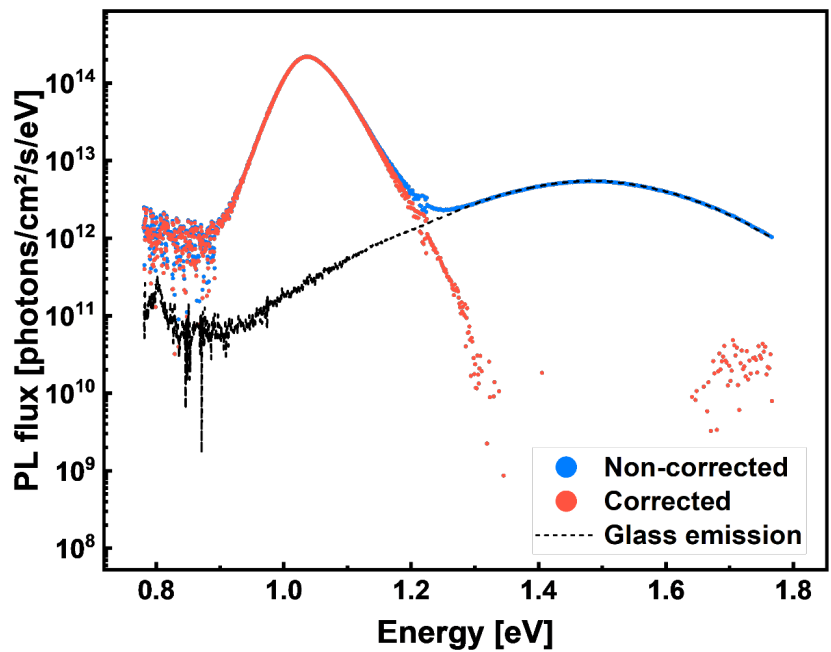

FIG. S1. Semi-log scale – absolute photoluminescence of the absorber displayed in Fig. 2. The blue spectrum is the non-corrected emission of the absorber affected by the glass emission (black dotted line). The red spectrum is the resulting corrected spectrum after the glass emission is subtracted.

## S2.   Band gap determination

There exist plenty of ways to determine the band gap of an absorber. We compare 4 different methods: reflection, PL, EQE and GDOES. Fig. S2 shows the band gap values obtained for absorbers with diverse the sulfur composition. The three first methods agree on the band gap minimum, considering that in general, the optical band gap extracted from PL is lower than the band gap of the material. [1,2]

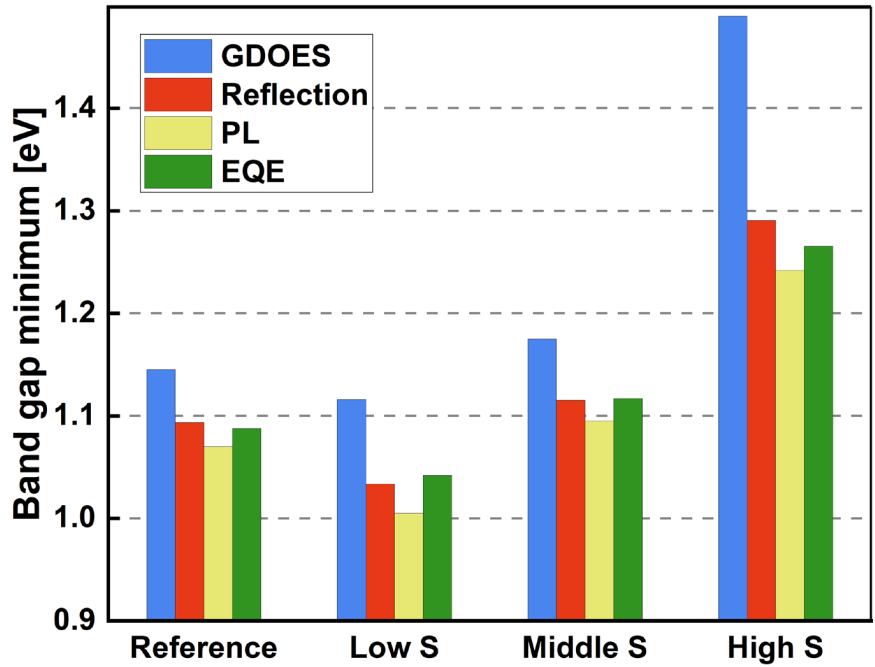

FIG. S2. Comparison of band gap minimum extracted from reflection, PL, EQE and GDOES for different absorbers of various S content.

## S3.   Low temperature photoluminescence

Photoluminescence was performed from the front side at low temperature (down to 10 K). It is observed that in the range 10-80 K, the high gap phase is also detectable. This supports the fact that some of the high gap phase material is located close to the front side.

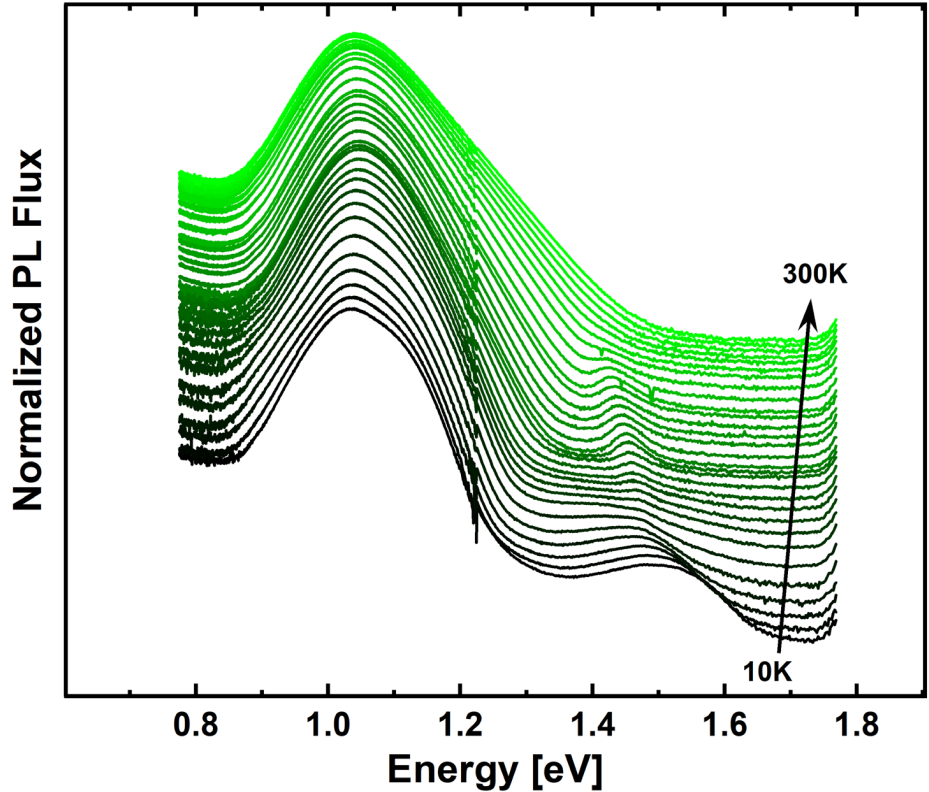

FIG. S3. Semi-log scale – Temperature dependent photoluminescence spectroscopy measurement from the front side. The data is normalized to the low energy peak and manually shifted for better display. In addition to the low energy peak, a second recombination channel is detected at high energy for temperatures between 10 K and 80 K.

## S4.    OVC phases from Raman spectroscopy

OVC phases were detected by means of Raman spectroscopy under 532 nm excitation wavelength. Using a laser of wavelength 785 nm, which is closer to the band gap of the OVC phases provides condition close to resonance and therefore more of the vibrational modes can be detected.

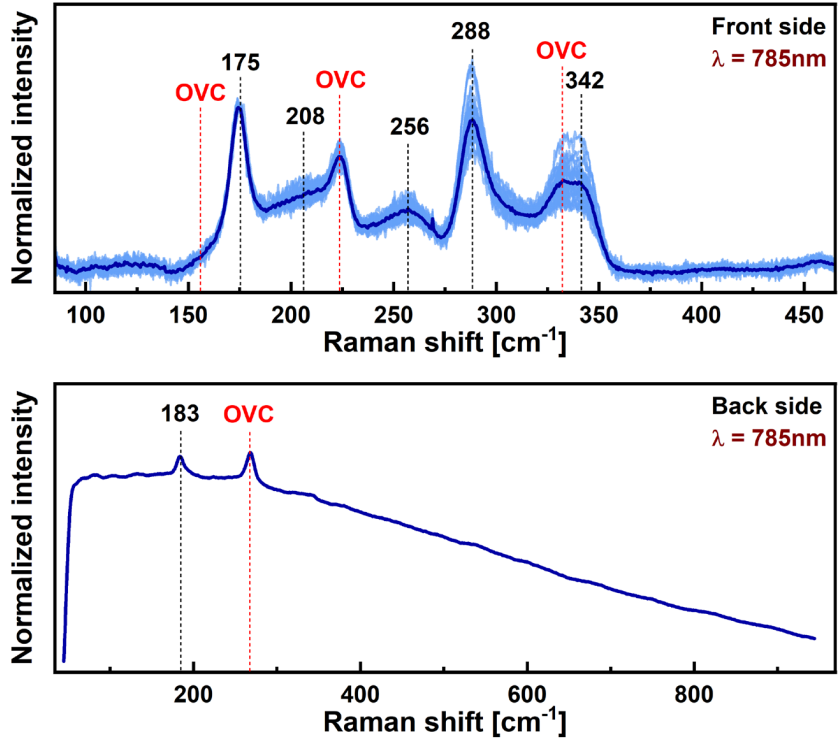

FIG. S4. Raman spectra measured from the front and back side under an excitation wavelength of 785 nm. From the back side, the background noise is quite high due to the photoluminescence signal under this excitation wavelength which prevents its correction by standard methods.

## S5. Cathodoluminescence of cross section

Cross sectional cathodoluminescence is performed on different positions of the absorber shown in Fig. 5 in the main text. The colormaps display the position of the maximum of the CL emission. In most of the cases (see Fig. S5), an abrupt transition between blue and red in observed, indicating that two distinct phases are coexisting, and no gradual band gap increase is detected.

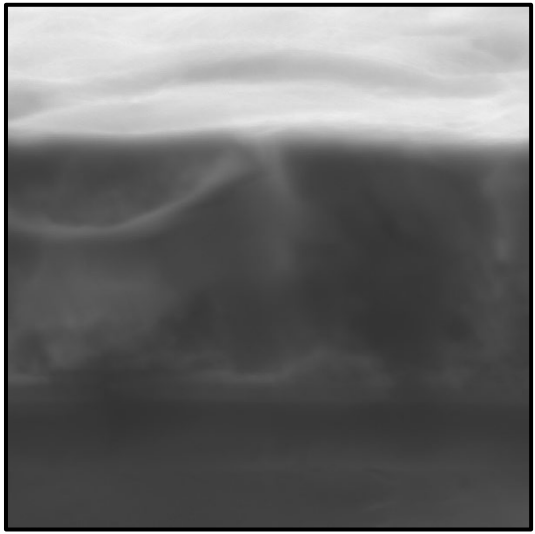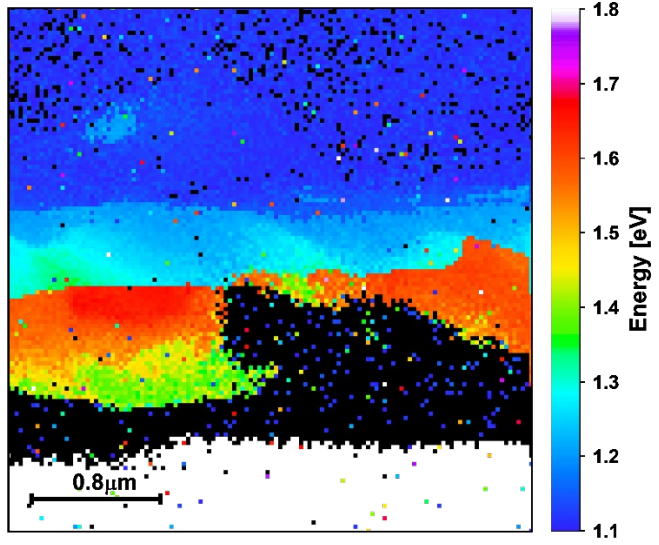

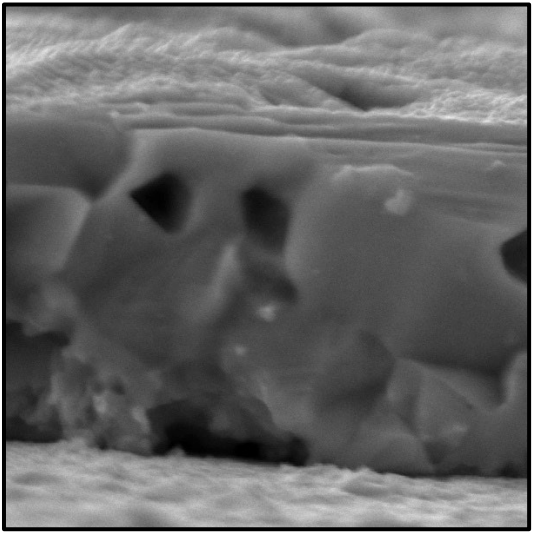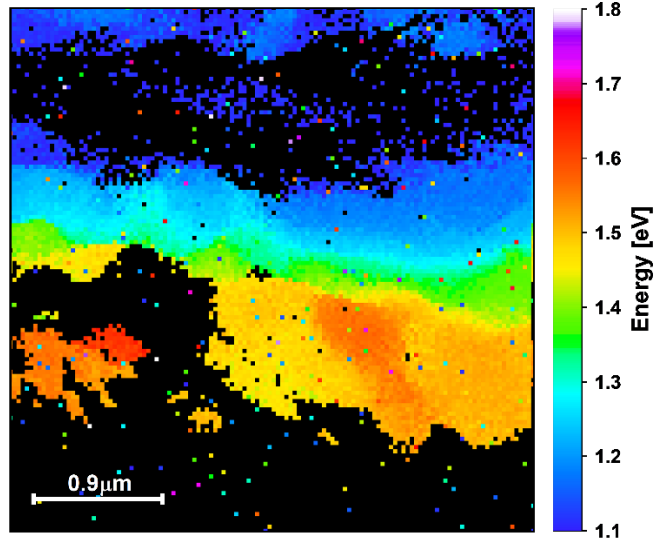

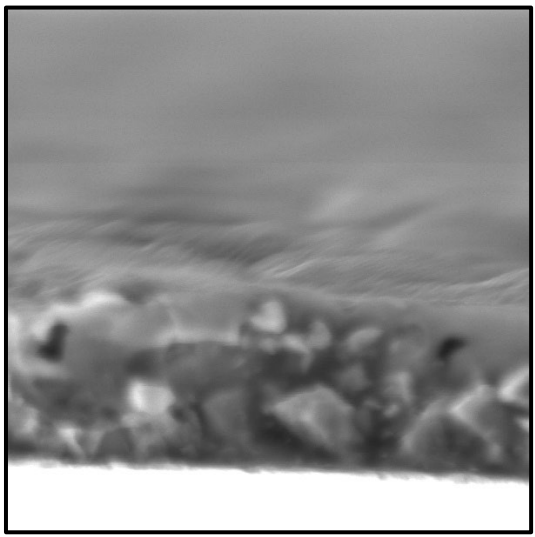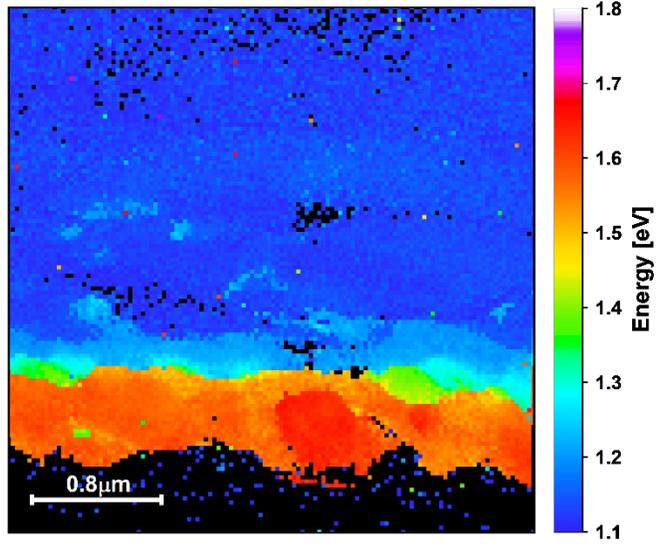

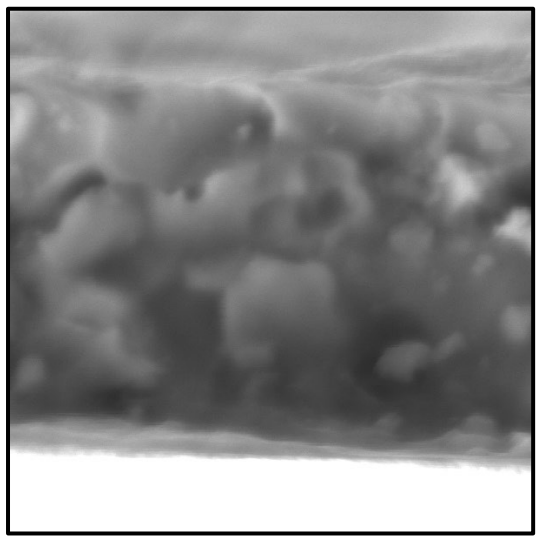
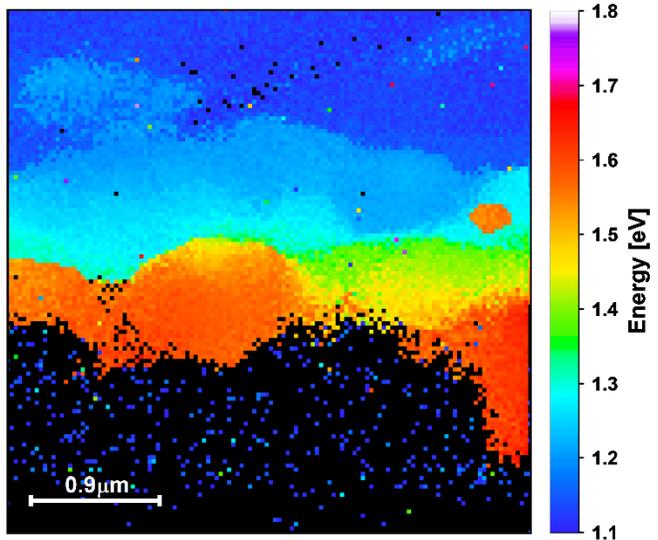

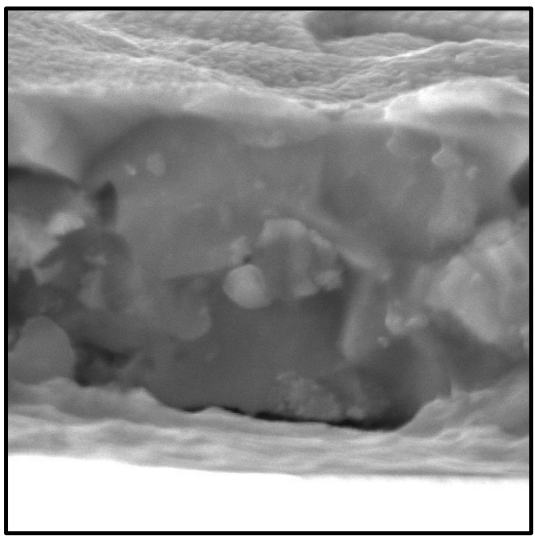
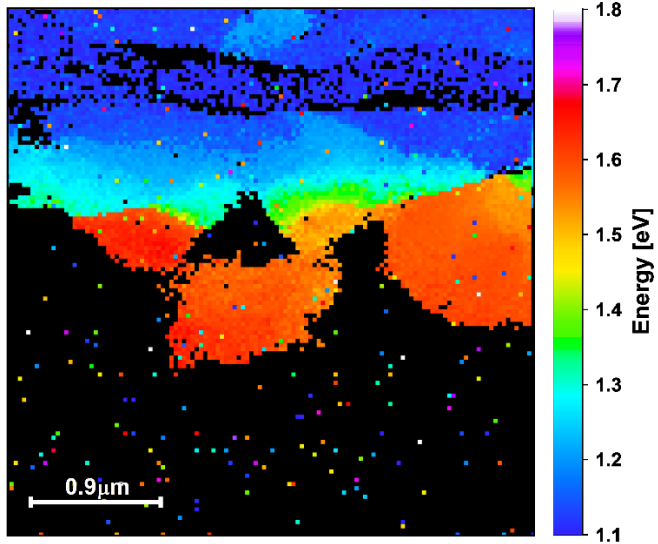

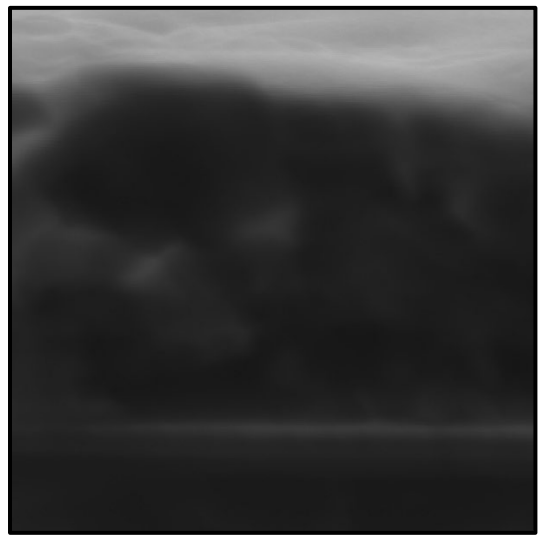
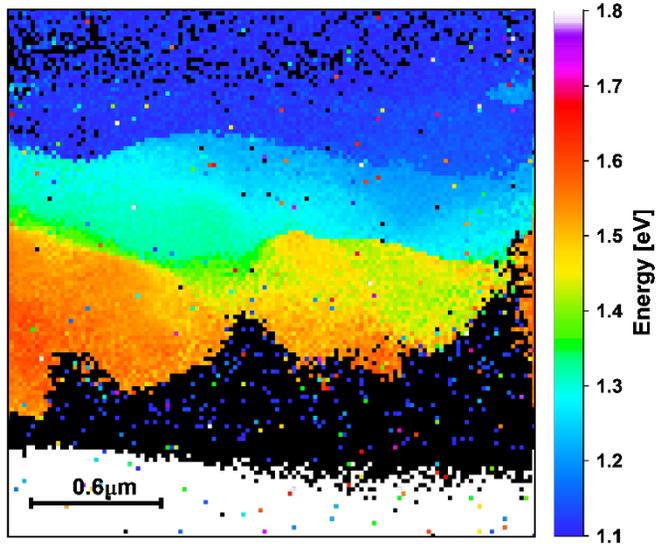

FIG. S5. Left: SEM cross section image of the region of interest where the CL is performed. Right: Corresponding colormap of the maximum of the CL emission. Orange-red corresponds to the high gap phase, green-yellow to the intermediate one and blue-cyan to the low gap phase. The white region at the bottom is molybdenum and the black is too noisy to identify peaks. The scale is identical for the right and left.

## S6.    Gradual etching from the back side

Samples from region III (see Fig. 1b) in the main text) have been gradually etched from the back side, as described in the main text. Fig. S6 shows the PL spectra at different final thicknesses of these two samples. Once the steep Ga step is etched away, an increase of the low energy PL peak is observed. The etching step is shorter in these cases than for the sample in Fig. 5. Therefore, it is clearer that the intensity increase is gradual; the more material is etched away, the more the intensity increases.

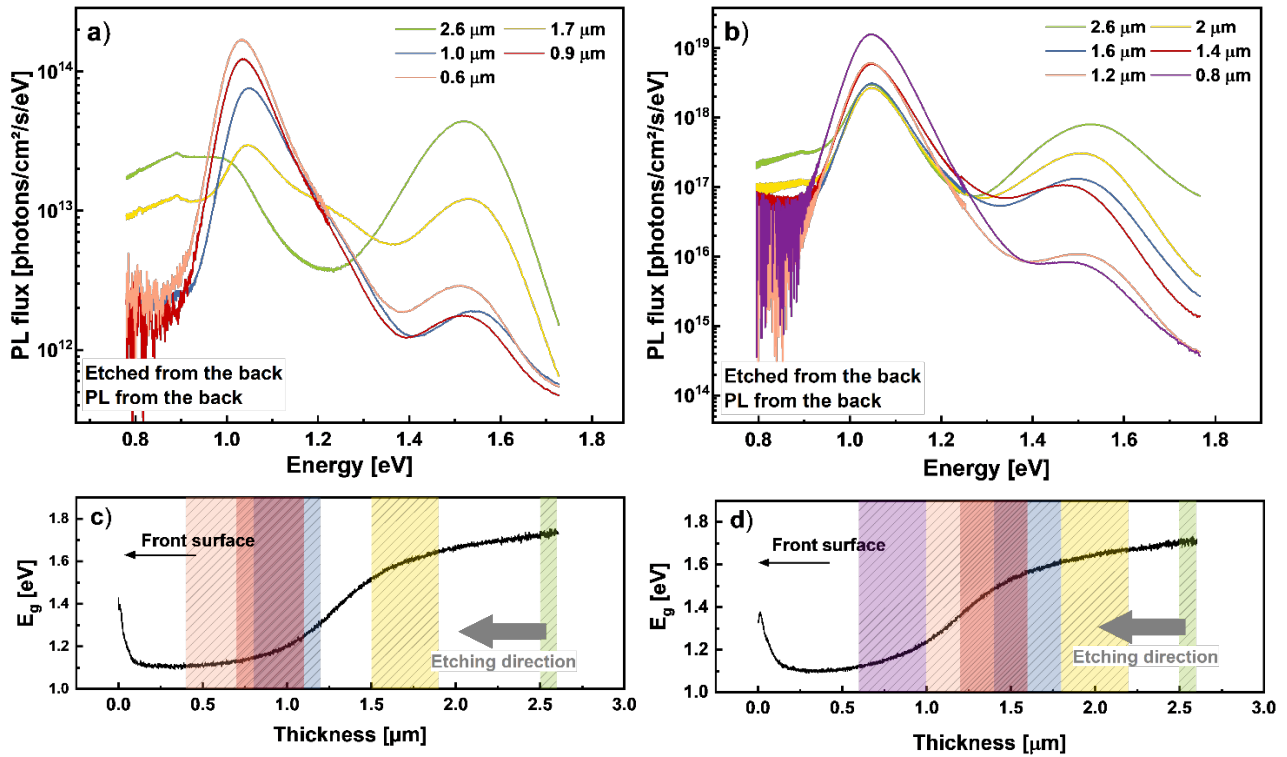

FIG. S6. Top: non-absolute photoluminescence spectra at different depths after the absorber is gradually etched from the back side for a) sample 17 and b) sample 13. The legend indicates the remaining thickness after the etching. Bottom: band gap profiles determined from GDOES. The colored bars indicate the etching depth, including errors.